\documentstyle[11pt]{article}
\begin{document}
\begin{flushright}
gr-qc/9508057\\
Alberta-Thy-18-95\\
\end{flushright}

\begin{center}
{\LARGE\bf Entropy of eternal black holes
\footnote{To appear in the proceedings of the Sixth Canadian Conference on
General Relativity and Relativistic Astrophysics, Fredericton, Canada, May
25-27, 1995}
}
\vskip .5cm
Erik A. Martinez
\footnote{Address after October 1st, 1995:
Center for Gravitational Physics and Geometry, \\
Department of Physics, The
Pennsylvania State University, University Park, PA 16802-6300, USA}
\vskip .5cm
{{\it Theoretical Physics Institute, University of Alberta\\
Edmonton, Alberta, Canada T6G 2J1}\\
{\tt martinez@phys.ualberta.ca}}

\end{center}
\vskip .3cm

\begin{abstract}
The entropy of a quantum-statistical system which is classically approximated
by a general stationary eternal black hole is studied by means of a
microcanonical functional integral.
This approach opens the possibility of including explicitly the
internal degrees of  freedom of a physical black hole in path integral
descriptions of its thermodynamical properties. If the functional integral is
interpreted as the density of states of the system, the corresponding entropy
equals  ${cal S} = A_H/4 - A_H/4 = 0$ in the semiclassical approximation,
where $A_H$
is the area of the black hole horizon. The functional integral reflects the
properties of a pure state.
\end{abstract}

\section{Introduction}

The relationship between the entropy of a physical black hole and its internal
degrees of freedom remains a subject of active research. A natural question to
ask in this regard is: can these degrees of freedom be effectively included in
a functional integral description of black hole entropy?
In an attempt to give an affirmative answer to this question, we investigate in
this paper a microcanonical functional integral when applied to a
quantum self-gravitating statistical system that includes spacetimes whose
topology and boundary conditions coincide with the ones of (either distorted or
Kerr-Newman) eternal black holes.

A proposal for the density of states of a gravitational
system defined in terms of a  microcanonical functional integral  has been
suggested recently in Ref. \cite{BrYo2}.
This integral is defined as a formal sum
over Lorentzian geometries. The black hole density of states is obtained
from this functional
integral  when the latter is approximated semiclassically by using a
complex metric whose boundary data at its single boundary surface coincide with
the boundary data of a Lorentzian, stationary black hole.
The density of states defined accordingly equals the exponential of
one fourth of the area of the black hole horizon.
This proposal opens the possibility of determining the
thermodynamical properties of black hole systems starting from a sum
over real Lorentzian geometries.
However, several problems remain in this approach.  First,  a spacelike
hypersurface $\Sigma$ that describes the initial data of a  Lorentzian black
hole  has to cross necessarily the event horizon and eventually intersect the
interior singularity.
Second, a  Lorentzian, stationary black hole is not a extremum of the
microcanonical action for a spacetime region with a single timelike
boundary surface.
This implies that the black hole density of states
whose boundary data correspond to the ones  of a   Lorentzian,
stationary black hole cannot be approximated semiclassically  by
using the same Lorentzian metric that motivates its boundary conditions.
It is therefore necessary to use complex metrics to evaluate the
Lorentzian functional integral in a
steepest descents approximation. This procedure yields the correct result
for the  entropy but conceals its origin: the interior of the Lorentzian
black hole literally disappears by virtue of this procedure, leaving
effectively only a periodically identified
Euclidean version of the  ``right" wedge region of a
Kruskal diagram. The properties of the black hole interior become encoded  in a
set of conditions at the so-called ``bolt" of the complex geometry.
In this approach the statistical origin of entropy and
its relationship to the
internal degrees of freedom of a black hole remain obscure.

The  problems mentioned above and the role of internal
degrees of freedom in  functional integral descriptions of black hole
thermodynamics can be addressed by explicitly
considering the  eternal version of a black hole \cite{Ma}.
The excitations of the physical black
hole can be associated with the deformations of an initial global Cauchy
surface $\Sigma$ of the eternal black hole.
In general, the spatial slices $\Sigma$ that foliate an
eternal black hole are (deformed) Einstein-Rosen bridges with wormhole topology
$R^1 \times S^2$. The spacetime is composed of two  wedges  $M_+$ and $M_-$
located in  the right  and left  sectors of a Kruskal diagram.
Internal and external degrees of freedom of the black hole can be
easily identified in this approach since the hypersurfaces $\Sigma$ are
naturally divided in two parts $\Sigma_+$ and $\Sigma_-$ by a bifurcation
two-surface $S_0$  where
the lapse function $N$ vanishes.
While the  ``external" degrees of freedom of the original black hole are
naturally given by the initial data at $\Sigma_+$,
its ``internal"  degrees of freedom can be identified with initial data defined
at $\Sigma_-$ \cite{FrMa2}.

\section{Microcanonical action and functional integral}

Consider a spacetime region whose three-dimensional
timelike boundary surface $B$ consists of
two disconnected parts $B_+$ and $B_-$.
The microcanonical action $S_m$ is the action appropriate to a variational
principle in which the fixed boundary conditions at the timelike  boundaries
$B_\pm$ are not the spacetime three-geometry
but the surface energy density
$\varepsilon$, surface momentum density $j_a$, and boundary two-metric
$\sigma_{ab}$ \cite{BrYo2,BrMaYo}.
The covariant form of the microcanonical action for a  general
spacetime whose timelike surfaces $B_+$ and $B_-$ are located in the
regions $M_+$ and $M_-$ respectively has been presented in Ref. \cite{Ma}. Its
Hamiltonian form  is easily obtained under a
$3+1$ spacetime split by recognizing that  there exists a direction of time at
the boundaries $B_\pm$ \cite{Ma}.
By introducing the  momentum $P^{ij}$ conjugate to the three-metric
$h_{ij}$ of $\Sigma$
for the so-called ``tilted" foliation \cite{FrMa2}
and integrating the kinetic part of the volume integral,  the action becomes
$S_m = \int_M d^4x [ P^{ij} \dot h_{ij} - N{\cal H} - V^i {{\cal H}_i}]$,
where the dot denotes
differentiation with respect to time, $V^i$ denotes the shift vector, and
the gravitational contributions to the Hamiltonian and momentum constraints are
given by the usual expressions.
Observe that the action vanishes identically for stationary solutions of
the vacuum Einstein equations describing stationary eternal black holes.
In this case $\dot h_{ij}=0$, the
constraint equations are satisfied, and no boundary  terms remain in the
action.

Consider now the microcanonical functional integral for a
gravitational system whose timelike boundary
surfaces $B_\pm$ are located in $M_\pm$. The path integral
\begin{equation}
{\bar {\nu}}[\varepsilon_+, j_+, \sigma_+ ;
\varepsilon_-, j_-, \sigma_-] = \sum_{\sl M} \int {\cal D}H \exp(i S_m)
\label{ournu}
\end{equation}
is a  functional of the  energy density $\varepsilon_\pm$,
 momentum density $j_\pm$, and two-metric $\sigma_{\pm}$ at the
boundaries $B_+$ and $B_-$.
The sum over $M$  refers to a sum over manifolds of different
topologies whose boundaries have topologies
$B_+ \equiv S_+ \times S^1 = S^2 \times S^1$ and
$B_- \equiv S_- \times S^1 = S^2 \times S^1$. The element $S^1$ is due to the
periodic identification in the global time direction at the boundaries when the
initial and final hypersurfaces are identified.   The  integral is a  sum over
periodic Lorentzian metrics that satisfy the boundary conditions at $B_+$ and
$B_-$.

The eternal black hole functional integral ${\bar {\nu}}_{*}$ is obtained
when the boundary data $(\varepsilon_+, j_+, \sigma_+)$
and  $(\varepsilon_-, j_-,
\sigma_-)$ of the geometries summed over correspond to the boundary data of a
general Lorentzian, stationary eternal black hole. The boundary
data  of this solution can be determined at $S_+$ and $S_-$ for each  slice
$\Sigma$. Observe that by virtue of the gravitational constraint equations,
these data determine
uniquely the size of the black hole horizon \cite{ensembles} and are such that
the two-metric is continuous at this horizon.

We evaluate now the functional integral in the semiclassical
approximation. This requires finding a four-metric that extremizes the
action $S_m$ and satisfies the boundary conditions  $(\varepsilon_+, j_+,
\sigma_+)$  at  $S_+$ and  $(\varepsilon_-, j_-, \sigma_-)$
at $S_-$. Since the classical Lorentzian eternal black hole metric can be
periodically identified  and placed on a
manifold  whose two spatial boundaries have the desired topologies,
the resulting metric can be used to approximate the path integral.
The periodic identification alters neither the constraint
equations nor the boundary data.
In the semiclassical approximation the functional integral ${\bar{\nu}}_{*}$
becomes \cite{Ma}
\begin{equation}
{\bar{\nu}}_{*}[\varepsilon_+,j_+, \sigma_+;  \varepsilon_- ,j_-,\sigma_-]
\approx   \exp \big( i S_m[{\tilde N}, {\tilde V}, {\tilde h}] \big)
\approx \exp \big( 0 \big)\ ,
\label{seminuL} \end{equation}
since the microcanonical action
$S_m[{\tilde N}, {\tilde V}, {\tilde h}]$
evaluated at the periodically identified geometry vanishes identically
if the stationarity condition and the constraints are satisfied.

It is illustrative to consider now a
complex  four-metric which also extremizes the microcanonical action  for
eternal black hole  boundary conditions and which can be used to reevaluate the
path integral (\ref{ournu}) in a steepest descent approximation.
This alternative approximation of the quantity ${\bar{\nu}}_{*}$ is useful in
understanding the relationship of the result (\ref{seminuL}) with the density
of states for an ordinary  (that is, non-eternal) black hole \cite{BrYo2}.
The complex metric can be obtained from the
Lorentzian eternal black hole metric  by  a complexification  map $\Psi$
defined
by  $\Psi(N)= -iN$,
$\Psi(V^i) = -iV^i$. This map   preserves the reflection symmetry and the
canonical variables $h_{ij}$ and $P^{ij}$ of the Lorentzian
solution.  The complex geometry consists of two complex sectors ${\bar M}_+$
and
${\bar M}_-$ which join at the locus of points at which the lapse vanishes.
This geometry is also a solution of Einstein equations if one requires that
conical singularities do not exist at that locus for every $\Sigma$.
To do this, it is necessary to puncture each complex sector and to close
smoothly the geometry at the inner boundaries
$^3\! H_\pm = ^2\! H_\pm \times S^1$ of ${\bar M}_\pm$,
where $^2\! H_{\pm}$ denotes the
intersection of the slices $\Sigma_{\pm}$ with the black hole horizon
for the Lorentzian metric \cite{Ma}. After imposing these regularity
conditions,
the topology of each sector ${\bar M}_\pm$  becomes $R^2 \times S^2$.
However, each element $^3\! H_+$ and $^3\! H_-$ does contribute a term to the
microcanonical action for the complex geometry.

The regular complex metric is not included in the sum over Lorentzian
geometries ${\bar{\nu}}_{*}$ in (\ref{ournu}) but can be used to
approximate it by distorting the contours of
integration for both lapse and shift
into the complex plane \cite{BrYo2}. In this approximation the path
integral becomes
${\bar{\nu}}_{*}
\approx  \exp ( i S_m[-i{\tilde N}, -i{\tilde V}, {\tilde h}])$,
where $S_m[-i{\tilde N}, -i{\tilde V}, {\tilde h}]$ is the
action  of the complex metic when the smoothness of the
geometries at  $^3 \! H_+$ and $^3 \! H_-$ is inforced.
This action turns out to be
$ S_m[-i{\tilde N}, -i{\tilde V}, {\tilde h}] = -{i} A_+/4
+{i} A_- /4$,
where $A_+$ and $A_-$ denote the surface area of the elements
${^2\!H_{\scriptscriptstyle +}}$ and  ${^2\! H_{\scriptscriptstyle -}}$
\cite{Ma}.
Since the periodic identification and the complexification $\Psi$ do not alter
the boundary data nor the gravitational constraint equations, the area $A_+$ of
${^2\!H_{\scriptscriptstyle +}}$ coincides with the area $A_-$ of
${^2\!H_{\scriptscriptstyle -}}$:
$A_+ (\varepsilon_+, j_+, \sigma_+) = A_- (\varepsilon_-, j_-, \sigma_-) \equiv
A_H$.
This implies that, in agreement with (\ref{seminuL}), the eternal black hole
functional integral is
${\bar{\nu}}_{*}[\varepsilon_+ ,j_+, \sigma_+; \varepsilon_-,
j_-, \sigma_-]  \approx \exp{( A_H/4 - A_H/4)} = \exp (0)$
in the ``zero-loop" approximation.

If the microcanonical functional integral (\ref{ournu})  is
interpreted as the density of states of the statistical system, it is possible
to express ${\bar{\nu}}_{*}$ approximately  as
${\bar{\nu}}_{*}[\varepsilon_+, j_+, \sigma_+; \varepsilon_-, j_-, \sigma_-]
\approx \exp({\cal S} [\varepsilon_+, j_+, \sigma_+; \varepsilon_-, j_-,
\sigma_-]) $,
where ${\cal S}$ represents the total entropy of the
system.  The above result
implies that  the entropy  for the system
in the semiclassical approximation is
\begin{equation}
{\cal S} \approx  {1\over 4}{A_H} - {1\over 4} {A_H}  = 0 \ ,
\label{entropy}
\end{equation}
where $A_H$ is the area of the horizon of the
physical eternal black hole solution that classically
approximates the system \cite{Ma}.
The total entropy is given formally by the subtraction
${\cal S} = {{\cal S}_+}[\varepsilon_+ ,j_+, \sigma_+]  - {{\cal
S}_-}[\varepsilon_- ,j_-, \sigma_-]$, where ${{\cal S}_+}$ and
${{\cal S}_-}$ can be interpreted as the semiclassical entropies
associated with
the external ($M_+$) and internal ($M_-$) regions respectively of the eternal
black hole system.

\section{Conclusions}

The functional integral (\ref{seminuL}) refers to a
quantum-statistical system  which is classically approximated by a general
stationary, eternal black hole solution of Einstein equations
within a  region bounded by two timelike surfaces $B_+$ and $B_-$.
Its semiclassical value  is a consequence of the choice of boundary data, the
gravitational
constraint equations, and the vanishing of the microcanonical action for the
four-geometries that satisfy  the boundary conditions and approximate the path
integral.   The calculation presented above applies to any distorted black hole
in the strong gravity regime.
It indicates  that a pure state (of zero
entropy) can be defined not only for
matter fields perturbations propagating in the spacetime of an eternal black
hole but also for the gravitational field itself. This is
physically appealing: the initial data for the eternal black hole specified at
the spacelike hypersurface $\Sigma$  contain all the information required for
the evolution of both the exterior and interior parts of a physical black hole.
The entropy associated with $\Sigma$ must therefore equal zero.
Since in a microcanonical description it seems natural to
relate  the external and internal degrees of freedom of a black hole with the
boundary data at the surfaces $B_+$ and $B_-$ respectively \cite{FrMa2}, we
believe that the microcanonical functional integral for eternal black hole
systems opens the possibility of extending path integral formulation of
black hole thermodynamics to situations when  internal degrees of freedom
are present and allows the study of gravitational
statistical properties in terms of a single pure state \cite{Ma}.
These conclusions are in complete agreement with thermofield dynamics
descriptions of quantum processes and, in particular, with the
application of this approach to black hole thermodynamics developed originally
by Israel \cite{Is} for small perturbations. They strongly suggest that
the thermofield dynamics description of quantum field processes in a curved
background can be extended beyond perturbations to the gravitational field
itself of distorted eternal black holes.

I am grateful to Valeri Frolov and Werner Israel for helpful comments.
This work was supported by the Natural Sciences and Engineering Research
Council of Canada.


\begin{thebibliography}{9}

\bibitem{BrYo2} J. D. Brown and J. W. York, Jr., Phys. Rev. {\bf D 47},
1420 (1993).

\bibitem{Ma} E. A. Martinez, Phys. Rev. {\bf D 51}, 5732 (1995).

\bibitem{FrMa2} V. Frolov and E. A. Martinez, ``Action and Hamiltonian
for eternal black holes", preprint, Alberta-Thy-32-94, gr-qc/9411001,
(1994).

\bibitem{BrMaYo} J. D. Brown, E. A. Martinez, and J. W. York, Jr., Phys.
Rev. Lett., {\bf 66}, 2281 (1991).

\bibitem{ensembles} J. D. Brown, G. L. Comer, E. A. Martinez, J. Melmed,
B. F. Whiting, and J. W. York, Class. Quantum Grav. {\bf 7}, 1433 (1990).

\bibitem{Is} W. Israel, Phys. Lett. {\bf 57A}, 107 (1976).

\end{thebibliography}
\end{document}